\documentclass[10pt,a4paper,twocolumn]{article}
\usepackage{sectsty,fullpage,authblk,amsmath,graphicx,bbding,nicefrac,adjustbox,upgreek,dblfloatfix,float}
\usepackage[allcolors=black,colorlinks=true]{hyperref}

\usepackage[T2A,T1]{fontenc}
\usepackage[russian,german,english]{babel}

\DeclareMathOperator*{\argmin}{arg\,min}

\sectionfont{\normalsize}
\subsectionfont{\normalsize}
\subsubsectionfont{\normalsize}
\paragraphfont{\normalsize}
\usepackage[font=small,labelfont=bf]{caption}

\let\oldsection\section
\renewcommand{\section}[1]{\vspace{-.75em}\oldsection{\!\!\!#1}\vspace{-.75em}}

\let\oldsubsection\subsection
\renewcommand{\subsection}[1]{\vspace{-.15em}\oldsubsection{\!\!\!#1}\vspace{-.25em}}

\title{Adaptive time-stepping for the Super-Droplet Method\\ Monte Carlo collision-coalescence scheme}
\author[1\,*\,\Envelope]{Emma Ware}
\author[2]{Piotr Bartman-Szwarc}
\author[1]{Adele L. Igel}
\author[3]{Sylwester Arabas}

\affil[1]{\small Department of Land, Air, and Water Resources, University of California Davis, Davis, CA, USA}
\affil[2]{\small Faculty of Mathematics and Computer Science, Jagiellonian University in Krakow, Poland}
\affil[3]{\small Faculty of Physics and Applied Computer Science, AGH University of Krakow, Poland}
\affil[*]{Research carried out while on Erasmus+ stay at the AGH University of Krakow, Poland}
\affil[\Envelope]{Correspondence to: ecware@ucdavis.edu}
\date{\vspace{-3em}}

\usepackage[backref=false,natbib,style=apa,backend=biber,sortcites=true,sorting=ynt,language=auto,autolang=other]{biblatex}
\DefineBibliographyStrings{english}{
  andothers = {et\addnbspace al\adddot}
}

\begin{document}
  \twocolumn[
  \begin{@twocolumnfalse}
    \maketitle
    \begin{abstract}

We~present an analysis of an adaptive time-stepping scheme for the Super-Droplet Method (SDM), a~Monte Carlo algorithm for simulating particle coagulation.
SDM represents cloud droplets as weighted superdroplets, enabling high-fidelity representations of microphysical processes such as~collision-coalescence.
However, the algorithm can undercount collisions when the expected number of~events is~not realizable given the superdroplet configuration, introducing a biased error referred here as the colllision deficit.
While SDM exhibits statistical spread inherent to Monte Carlo schemes, the deficit is a~systematic underestimation of~collision events.
This error can be addressed with adaptive time-stepping, which dynamically adjusts simulation time steps to~eliminate this deficit.
We~analyze the~behavior of~the deficit across a~wide range of timesteps, superdroplet counts, and initialization strategies, and explore trade-offs between accuracy and efficiency when adaptive time-stepping is implemented.
Using the classical Safranov-Golovin test case, we~show that the deficit increases with timestep and superdroplet count, and that adaptive time-stepping effectively removes the associated error without significant cost.
We~test a smooth continuum of initial distributions with extrema representing two different initialization methods, and find that while the deficit is sensitive to the choice of attribute-space sampling strategies, adaptive time-stepping substantially reduces the difference, allowing for users to~choose initialization methods optimized for other processes. We also propose a method of visualization, capturing both the attribute sampling, droplet interactions over multiple timesteps, and the deficit using network connectivity graphs.
In 2-D flow-coupled simulations, we find the deficit can have a stronger effect on convergence than previously shown, with uncorrected deficit delaying the onset of precipitation.
While we focus on liquid-phase clouds, the adaptive time-stepping scheme for SDM is also applicable to ice- and mixed-phase systems.
We strongly recommend implementation of~adaptive time-stepping, as it efficiently eliminates coalescence errors, increases robustness to initialization choices, and can have a noticeable impact in bulk properties in multi-cell simulations, with implications for large-scale cloud and precipitation modeling.

    \end{abstract}
  \end{@twocolumnfalse}
  \vspace{.5em}
]    

\section{Introduction}\label{sec:intro}

The Super-droplet Method (SDM), introduced in~\citet{Shima_et_al_2009}, is a Monte Carlo approach for modeling collisional growth in cloud microphysics.
It offers a computationally feasible way of representing collisional growth within particle-based descriptions of cloud microphysics, in which a Lagrangian framework is used to track the evolution of computational particles represented by a~set of~properties, such as size, composition, and location in space.
Lagrangian models are a natural choice for representing the dispersed phase of colloids, and have long been identified as suitable for simulations of cloud droplet condensational growth, free of numerical diffusion in~both size and space 
\citep[see][for the seminal box- and flow-coupled Lagrangian model applications, respectively]
{Howell_1949,Clark_and_Hall_1979}.
They allow for a uniform treatment of particles over the full span of~the size spectrum, without arbitrary categorization into aerosol, cloud and rain subpopulations, a known limitation of models with ``bulk'' \citep{Igel_et_al_2022} as well as ``bin'' \citep{Witte_et_al_2022} microphysical schemes.
This helps accurately model the diffusional and collisional evolution of cloud droplet size and composition. 

In warm clouds, the process responsible for growth of droplets to sizes large enough to~precipitate is collision-coalescence \citep[see e.g][for an overview of the state of cloud modeling and represented processes]{Grabowski_et_al_2019}.
Coagulation and coarsening models have been the focus of diverse and sustained research interest, with applications spanning fields from material science, colloidal chemistry and combustion, flock dynamics in~biology, genetics and genealogy, to aerosols (including atmospheric clouds and precipitation) to~astrophysics \citep[see][for a multidisciplinary collection of works]{Bertoin_et_al_2007}.
The~classical deterministic model of binary coagulation was introduced by \citet{Smoluchowski_1917}.
Monte Carlo methods offer a~wide range of algorithmic approaches suitable for simulating stochastic coagulation, in~which discrete events and rare fluctuations play a key role \citep[see, e.g., discussion in][]{Gillespie_1975,Dziekan_and_Pawlowska_2017,Grabowski_et_al_2019}. 
Among the particle-resolved Monte Carlo algorithms for~collisional dynamics in aerosol systems (clouds included), there are the seminal methods of \citet{Gillespie_1975} and \citet{ORourke_1981}, 
the PartMC algorithms (based on the Smoluchowski Coagulation Equation) \citep[][]{Riemer_et_al_2009,DeVille_et_al_2011,DeVille_et_al_2019} and SDM (based on the underlying collision process) \citep[referred to as the All-Or-Nothing AON scheme in][]{Unterstrasser_et_al_2017}.

\begin{figure}[ht!]
    \centering
      \includegraphics[width=
      1.02\linewidth]{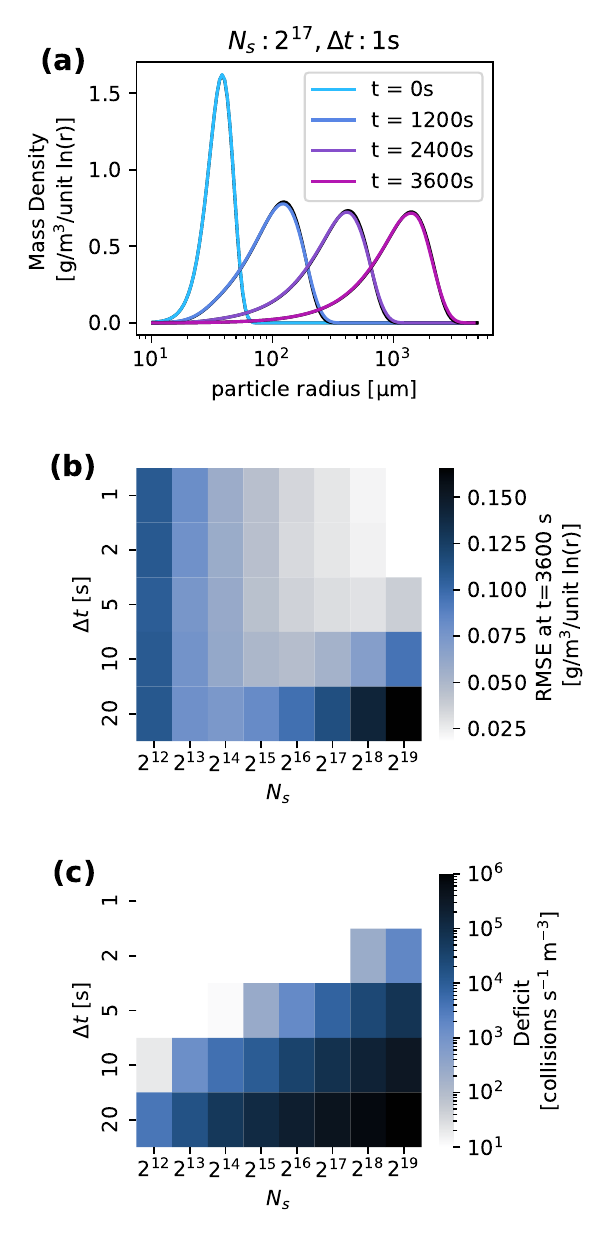}
      \vspace{-3em}
    \caption{Results from an ensemble of 50 Monte Carlo box-model simulation using the additive kernel -- as in~Fig.~2a in \citet{Shima_et_al_2009}: (a)~size spectrum at~four different instances with colored curves corresponding to SDM results ($N_s=2^{17}$ superdroplets, $\Delta t=5$s, $\Delta V=10^6$m\textsuperscript{3}, mass concentration of $1$g\,m\textsuperscript{-3}, kernel parameter $b=1500$s\textsuperscript{-1}, initial concentration of~2\textsuperscript{23}m\textsuperscript{-3} particles) and black curve denoting the analytic solution; (b)~RMSE of~the SDM vs. analytic solution at~final timestep for different spectral and temporal resolutions; (c)~average collision deficit per time and volume for the same set of~simulations as in panel~(b).}
    \label{fig:Shima_2009_deficit}
\end{figure}

SDM has received substantial attention within~the atmospheric cloud modeling community \citep[see, e.g., the review of][]{Morrison_et_al_2020}, and 
is~noted to be a particularly accurate approach in~terms of convergence to analytical solutions \citep{Unterstrasser_et_al_2017}.
In SDM, computational particles, or superdroplets, represent integer multiplicities of real droplets that evolve through physical processes.
The SDM algorithm avoids numerical diffusion, maintains constant state vector size (conservation of~the~number of~simulation particles during collisions), obeys strict mass conservation, has linear computational cost scaling with the state vector size, and is~embarrassingly parallel. 

However, recent work has shown that SDM can underestimate collision rates under certain conditions; we hereinafter refer to the number of omitted collisions in the numerical solution as the deficit.
The deficit was defined and quantified in~\citet{Bartman_AMS_2023}, building on earlier observations reported in \citet{Dziekan_and_Pawlowska_2017, Unterstrasser_et_al_2020}.
It arises when the expected number of droplet collisions between a superdroplet pair exceeds what is realizable given the superdroplet multiplicities. 
In~such cases, the~number of~experienced collisions is capped accordingly, leading to the systematic underestimation of~the~number of collision events.
\citet{Unterstrasser_et_al_2020} shows a case where the limit is reached within~the simulation, but the convergence of~the simulation is~not affected (Sec. 3.2, Fig. 8 therein). 
This study aims to characterize when and where the deficit becomes significant, how it~scales with key simulation parameters, and how it can be entirely avoided using adaptive timestepping.

To define the deficit, we embrace the notation used in \citet[][section 4.1.4 and 5.1.3]{Shima_et_al_2009}.
SDM allows for multiple collision events within a~timestep. 
The integer number of superdroplet collision events per timestep for a given pair is~denoted as $\gamma_{jk}$ (where $j$~denotes the ``donor'' and $k$~denotes the ``collector'' droplet).

Pairwise probabilities greater than unity result in multiple collisions, and these can stem from large timesteps, high droplet number concentrations, or from scaling the probabilities due to the
linear-in-$N_\text{s}$ pair sampling technique employed by SDM, where a~non-overlapping set of~pairs is chosen to test in~a~given timestep for events (testing all possible pairs incurs quadratic-in-$N_\text{s}$ expense).

During an event, the multiplicity of~the collector (${\xi}_k$) determines how many droplets move from the donor superdroplet to the collector.
If~the resulting number of collisions from the predicted events exceeds the donor’s available multiplicity (${\xi}_j$), the algorithm caps the~number of~events to~$\lfloor{\xi}_{k}/{\xi}_{j}\rfloor$, now referred to as $\tilde{\gamma}_{jk}$.
This ensures conservation of mass and superdroplet number, but results in~some collisions being omitted. 
This is~the collision deficit, and is defined as the difference between the expected number of~collision events and the realizable number of~collision events, scaled by the multiplicity of~the collector superdroplet:
\begin{equation}
\label{eq:deficit}
    \text{deficit} \equiv \xi_k(\gamma_{jk} - \tilde{\gamma}_{jk})\text{~.}
\end{equation}
Allowing multiple collisions is a defendable tradeoff for the computational efficiency of a~longer timestep or a parallelizable subset of~pairs tested. Although a~longer timestep reduces accuracy of~the simulation, the effects are stochastic and do not favor overpredicting or underpredicting collisions.
However, the resulting bias from capping ${\gamma_{jk}}$ at $\lfloor{\xi}_j/{\xi}_k \rfloor$ monotonically underpredicts the~number of collisions and hence the average size of particles in~the system in a timestep \citep[see p. 13512 in][]{Dziekan_and_Pawlowska_2017}.
Figure~\ref{fig:Shima_2009_deficit} illustrates where the deficit manifests in practice.
Panel (a) recreates Fig. 2a of \citep{Shima_et_al_2009} with the original settings, presenting the evolution of an initially exponential droplet size distribution subject to collisionl growth with an additive collision kernel, comparing to the  Safranov-Golovin analytical solution \citep{Safranov_1962,Golovin_1963}; panel (b) maps the root-mean-square error (RMSE) (with respect to the analytical solution) at the final timestep across a~range of~superdroplet counts and timestep lengths,  $N_s$ and $\Delta t$; and panel (c) shows the corresponding magnitude of~the collision deficit.
The figure shows that the deficit becomes significant in specific regions of~the parameter space.
The deficit is a function of~the timestep, but it also increases with larger $N_s$, leading to a~counterintuitive result that in regimes with deficit, increasing $N_s$ (i.e., higher size spectral resolution) actually \textit{increases} the error (discussed in section \ref{sec:box-results})!

Rather than allowing for multiple events in~a~timestep, other particle-based models have introduced adaptivity to handle the issue when collision rates imply probabilities greater than one.
In \citet{Unterstrasser_et_al_2017}, an adaptive time-stepping scheme is applied in~the Remapping Algorithm \citep[RMA,][]{Andrejczuk_et_al_2010} to avoid negative multiplicities.
The Weighted Flow Algorithm \citep[WFA,][]{DeVille_et_al_2011}
adapts the~number of~candidate pairs based on expected collision rates, leveraging the fact that particle weights are continuous functions of droplet attributes rather than fixed multiplicities. 
In \citet{Dziekan_and_Pawlowska_2017},
an adaptive substep was also employed when exploring coagulation of droplets with multiplicties of 1, where multiple collections would not be~possible.
These precedents support the implementation of an adaptive approach within SDM that allows multiple collections but avoids the deficit, tailored to the specific formulation and multiplicity structure of SDM.

In \citet{Bartman_AMS_2023, deJong_et_al_2023}, adaptive time-stepping is included in~the development of PySDM, an open-source cloud microphysics package, as a solution to~the deficit.
This approach calculates the maximum allowable timestep that avoids deficit for all tested superdroplet pairs in a collision volume.
Substepping is triggered only when this condition is violated, ensuring that computational resources are spent only where the deficit would otherwise distort results. 
Using PySDM, we look at how implementing adaptive time-stepping improves accuracy across a~wide range of attribute initializations, superdroplet counts, and test cases, including both zero-dimensional box-model, and two-dimensional flow-coupled configurations.

\noindent
The remainder of the paper is organized as follows:\\
-- {\bf Section \ref{sec:methods}} lays out the original formulation of~the SDM collision step, and introduces the adaptive time-stepping formulation, explaining how pairwise constraints guide substep calculation. In this section we also discuss how attribute initialization strategies affect the deficit.\\
-- 
{\bf Section \ref{sec:box}} presents the results of~adaptive time-stepping in~the box model runs, looking at~a~large range of temporal and spectral resolutions and different initial spectrum sampling methods.\\
-- {\bf Section \ref{sec:flow}} provides an analysis of~the adaptivity scheme in a 2-D flow-coupled model and details a~load-balancing strategy for multi-cell scenarios in~parallel architectures.\\
-- {\bf Section \ref{sec:conclusions}} concludes with a recap of~the main points and a recommendation for implementing adaptive time-stepping.\\
-- {\bf Appendix \ref{sec:network}} provides a supplementary visualization of the droplet collisional interactions as~a~network graph. This highlights the relationship between the attribute sampling methods discussed in section~\ref{sec:methods} and the deficit.

\section{Methods}\label{sec:methods}

\begin{figure*}[ht!]
    \centering
    \includegraphics[width=\linewidth]{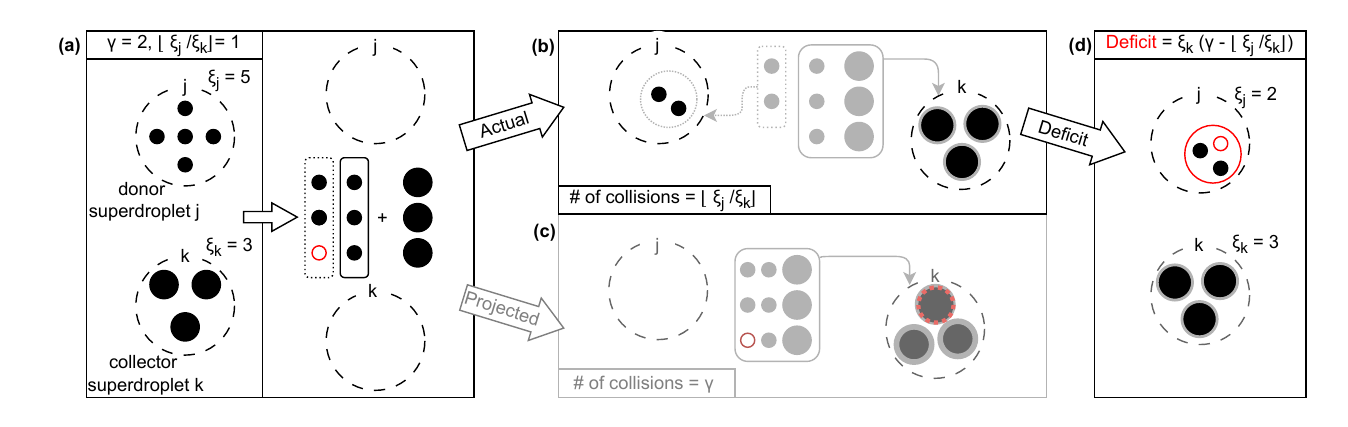}
    \vspace{-.5em}
    \caption{A schematic outlining the scenario in which a collision event between a donor superdroplet with multiplicity $\xi_j=5$ and a collector superdroplet with multiplicity $\xi_k=3$ leads to collision deficit. Panel~(a)~depicts the initial state, with  $\gamma_{jk}=2$ implying two collision events predicted; panel (b) depicts the actual outcome when the realizable number of collisions is limited by $\lfloor{\xi_j}/{\xi_k}\rfloor=1$; panel (c) shows the two projected events, out~of~which only one is realizable. This difference leads to the deficit $\xi_k \cdot(\gamma_{jk} - \lfloor{\xi_j}/{\xi_k}\rfloor) = 3 \cdot (2-1)=3$ shown in panel (d).
}
    \label{fig:deficit_problem}
    \vspace{.5em}
\end{figure*}

\subsection{Adaptive timestep nullifying the deficit}

The goal of~the adaptivity scheme is to prevent the~number of predicted collision events between a~pair of~superdroplets from exceeding what is~numerically representable, while still allowing for multiple collections.
Specifically, for any pair $\alpha$, we avoid the situation where $\gamma_{jk} > \lfloor\xi_{j} / \xi_{k}\rfloor$.
To~achieve this, we compute a maximum allowable substep length for each candidate pair.
The final timestep used in~the simulation cell is~then set to the smallest of~these pairwise limits, ensuring that all predicted collisions can be represented.

The pairwise probability $p_{jk}$ for two superdroplets $j$ and $k$ is defined as the probability of~two droplets colliding in a timestep and volume,
multiplied by the maximum multiplicty ($\xi$) of~the pair.
This is scaled by the ratio of the~number of pairs over the~number of non-overlapping pairs tested.  
$K$~is a collision kernel that takes in~droplet attributes such as size or terminal velocity, $N_s$~is~the~number of superdroplets in~the collision volume $\Delta V$, and $\Delta t$~is~the timestep.
\begin{equation}
    p_{jk} = { K({j}, {k}) \max(\xi_{j} , \xi_{k}) \frac{(N_s^2 - N_s)/2}{\lfloor N_s/2 \rfloor} }\frac{\Delta t}{{\Delta V}}\text{~.}
\end{equation}
The integer number of successful events $\gamma_{jk}$ is computed using a randomly drawn number $\phi$ from 0~to~1, with 
\begin{equation}
    \gamma_{jk} = \lceil p_{jk}- \phi \rceil
\end{equation}
\citep[equivalent to the equation calculating $\gamma_\alpha$ in algorithm step three in section 5.1.3 of][]{Shima_et_al_2009}.
In the original formulation, here is where $\tilde{\gamma}_{jk}$ referenced in the deficit definition (\ref{eq:deficit}) is calculated as 
\begin{equation}
\tilde\gamma_{jk} = \min(\gamma_{jk},\lfloor\xi_{j} / 
\xi_{k}\rfloor)
\end{equation}
which is needed to make the outcome of a collision event numerically realizable.
This is what can lead to the discrepancy between projected and actual outcomes, shown schematically in Fig.~\ref{fig:deficit_problem}.

Alternatively, we propose taking multiple substeps within a timestep so that $\gamma_{jk}=\tilde{
\gamma}_{jk}$.
We~substitute $\lfloor\xi_{j} / \xi_{k}\rfloor$ for 
$\gamma_{jk}$ to find the maximum allowed substep length:
\begin{equation}
    \Delta t_{jk}^{max} = \frac{\lfloor\xi_{j} / \xi_{k}\rfloor}{{\gamma}_{jk}}\text{~.} 
\end{equation}
For $\Delta t_{jk}\leq \Delta t_{jk}^{max}$, we have $\gamma_{jk} \Delta t_{jk} \leq \lfloor\xi_{j} / \xi_{k}\rfloor$ so~$\gamma_{jk} \leq \lfloor\xi_{j} / \xi_{k}\rfloor$ and the deficit for a single pair vanishes. 
The cell substep is defined by the smallest limiting timestep of~all pairs in~the cell:
        \begin{equation}\label{eq:substep}
            \Delta t_\text{substep} = \min \left(\min\limits_{{jk} \in \text{cell}} \Delta t_{jk}^{\max}, \Delta t \right)\text{~.}
        \end{equation}
Adaptively substepping when hitting this limit allows us to retain~the computational benefits of~a~longer time step whenever possible, while avoiding systematic underestimation of collisions. 

The attribute update for all substeps is performed as in the original formulation:
\begin{description}
  \item [(a)] If $\xi_{j} - \tilde{\gamma}_{jk}\xi_{k} > 0$
    \begin{equation}
      \begin{split} \label{eq:coal_attr_0}
        \xi_{j}' &= \xi_{j} - \tilde{\gamma}_{jk}\xi_{k} \\
        M_{j}' &=  M_{j}
      \end{split}
    \quad\quad
      \begin{split}
        \xi_{k}' &=\xi_{k} \\
        M_{k}' &=  M_{k} + \tilde{\gamma}_{jk} M_{j}
      \end{split}
    \end{equation}
  \item [(b)] If $\xi_{j} - \tilde{\gamma}_{jk}\xi_{k} = 0$
    \begin{equation}
      \begin{split}
        \xi_{j}' &= \lfloor\xi_{k}/2\rfloor \\
         M_{j}' &= \hat M_{k}
      \end{split}
    \quad\quad
      \begin{split}
        \xi_{k}' &= \xi_{k} - \lfloor\xi_{k}/2\rfloor \\
        M_{k}' &=  M_{k} + \tilde{\gamma}_{jk} M_{j}
      \end{split} 
      \label{eq:coal}
    \end{equation}
\end{description}
where M denotes any extensive attribute (e.g. mass, volume), and the primes represent the droplet state post collision event.

\subsection{Attribute sampling and collision deficit}\label{sec:alpha}

The accuracy and convergence behavior of SDM have been shown to be strongly influenced by the sampling methods involved in initialization of~superdroplet attributes \citep{Dziekan_and_Pawlowska_2017, Unterstrasser_et_al_2017}.
This initialization also affects the deficit, and thus employment of the adaptive time-stepping improves robustness with respect to the choice of sampling method.

With a given number distribution $n(a)$ and a proposed superdroplet distribution $p(a)$ over droplet attribute $a$, the resulting multiplicity function follows: $\xi(a) = n(a)/N_sp(a)$ \citep[][Sec.~5.3]{Shima_et_al_2020}.
Commonly explored superdroplet distributions include (but are not limited to):
\begin{itemize}
    \item[--] constant-multiplicity: $\xi(a)$ is~a~constant function, and  $p \propto n$;
    \item[--] uniform-in-$\log(r)$: $p(\log r)$ is~a~constant function, $\xi(\log r) \propto n(logr)$ and $r$ denotes radius;
    \item[--] uniform-in-radius: $p(r)$ is~a~constant function, and $\xi(r) \propto n(r)$.
\end{itemize}

For all of these methods, sampling can be performed both using a uniform random set of numbers, or deterministically (with equally spaced quantiles, radii, or log radii). 
Note that even with deterministic sampling of the initial attributes, SDM involves Monte Carlo pair selection and event triggering.
 Uniform-in-$\log(r)$ and uniform-in-$r$ initialization resolves the tail of~the probability distribution better than the constant multiplicity weighting, which focuses the resolution on the bulk of~the population.
The constant-multiplicity sampling is~also referred to as unweighted, quantile or~inverse-CDF sampling. 
Resolving the tail is~important for processes like rain formation, where large droplets are rare but are likely to trigger the onset of precipitation \citep[see, e.g.][]{Li_et_al_2023}.
However, the realization spread in~terms of microphysical variables in cases where the size spectrum is~primarily shaped by non-collisional processes, was reported to be smaller (i.e. higher accuracy) for constant multiplicity sampling \citep[][sec.~4.3]{Matsushima_et_al_2023}.

Algorithmically, SDM does best with a dynamic range of multiplicities, requiring smaller $N_s$ for convergence with respect to collision-coalescence analytic solutions \citep{Matsushima_et_al_2023,Dziekan_and_Pawlowska_2017,Unterstrasser_et_al_2017}.
This is because the dynamic range raises the average ${\xi}_j$ across all the~pairs in a given timestep (choosing~${\xi}_j$ to~be~the higher multiplicity) increasing the average pairwise collision probability $p_{\alpha}$ and spreading out the expected droplet collisions over more superdroplet pairs (see also discussion of~appendix Fig.~\ref{fig:network}).
This gets better coverage for the same amount of~pairs tested.
Not only does a dynamic range of~multiplicity lower the~number of~superdroplets required for convergence, but it also decreases the deficit, as~the limiting ratio ${\xi}_j/{\xi}_k$ is~on~average larger (again, choosing ${\xi}_j > {\xi}_k$).
However, the state changes dynamically throughout a~simulation as droplets undergo collisions, so~the benefits of~efficient initialization may may not persist in the long term.

For a~wider range of initialization options, \citet{Matsushima_et_al_2023} proposes a parameterization of the initialization of attributes, 
defining a~continuum of~distributions between two endpoints (such as evenly weighted constant-multiplicity and uniform-in-log($r$) sampling).
The resulting distribution $a$ is expressed as the $\alpha$-weighted Fréchet mean of~the two endpoint distributions $b_1$ and $b_2$, with $\mathcal{L}$ as a distance metric, and $a,b_1,b_2 \in C_k$, the possible set of discretized probability distributions:
\begin{equation}
    a = \argmin((1-\alpha) \mathcal{L}(a,b_1)+\alpha \mathcal{L}(a,b_2)) \text{~.}
\end{equation}
In one dimension (single attribute such as particle size), using the Wasserstein distance $W_2$ as $\mathcal{L}$, the above equation has a solution \citep{McCann_1997} which can be~found using displacement interpolation (the equation can be~used also in higher dimensional attribute spaces, but there is no analytical solution):
\begin{equation}
    F_a^{-1}(y) = (1-\alpha)F_{b_1}^{-1}(y) + \alpha F_{b_2}^{-1}(y) \text{~.} 
\end{equation}

Parameterizing the initialization
allows for flexible optimization among superdroplet distributions which may be more effective for modeling certain physical processes.
Using this alpha parameterization also allows us to look at the deficit across a continuous spectrum of distributions rather than discrete endpoints.

\begin{figure*}[ht!]
    \centering
    \includegraphics[width=\linewidth]{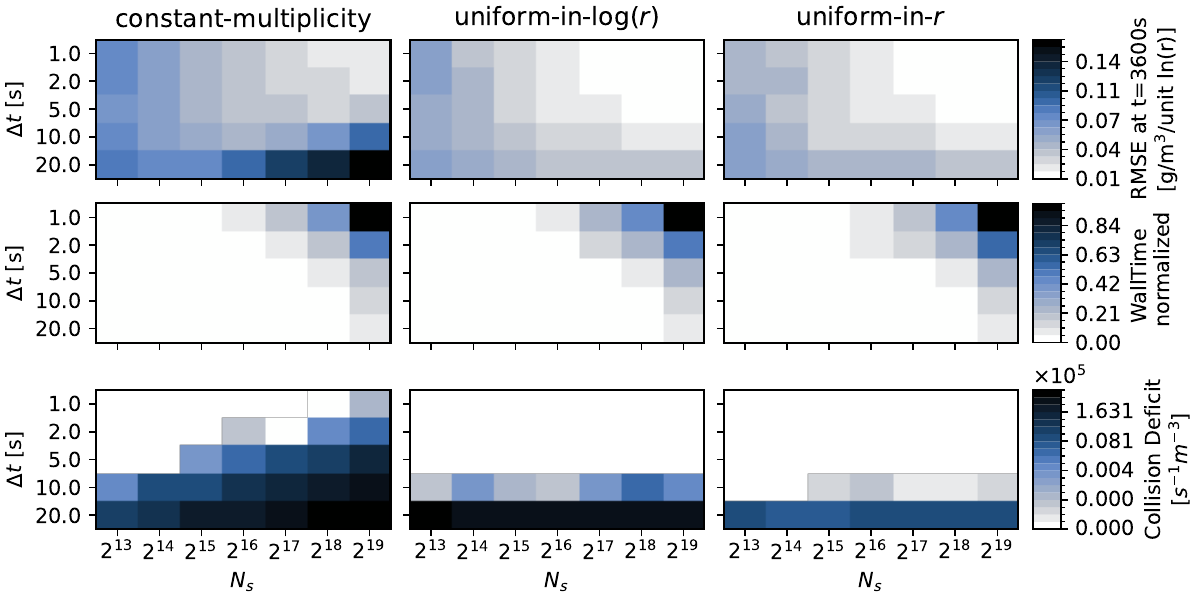}
    \vspace{-1em}
    \caption{
      Summary of box-model additive-kernel simulations without adaptivity compared against analytic solution.
      Columns group results for a particular initial spectrum sampling scheme.
      Diagrams in~the~top row present RMSE values; the middle row presents the wall time normalized as a fraction of the longest ensemble mean time (across attribute sampling methods); the bottom row presents average collision deficit.
      Each diagram cell represents a mean value for an~ensemble of~30~realizations.
      See~Sec.~\ref{sec:box-results} for discussion.
      }
    \label{fig:heatmaps}
    \vspace{.5em}
\end{figure*}
\subsection{Implementation}

For the implementation and analysis of~the deficit and adaptive time-stepping in~both the box model and the 2-D case, we use PySDM, an open-source just-in-time compiled Python package for modeling cloud microphysics \citep{Bartman_et_al_2022,deJong_et_al_2023}. 
PySDM includes multi-core CPU (threading) and GPU (CUDA) backends featuring the SDM adaptivity scheme.
The SDM adaptivity logic is also implemented in~the open-source Droplets.jl Julia package.

\section{Box-model simulations}\label{sec:box}

\begin{figure*}[ht]
    \centering
    \includegraphics[width=\linewidth]{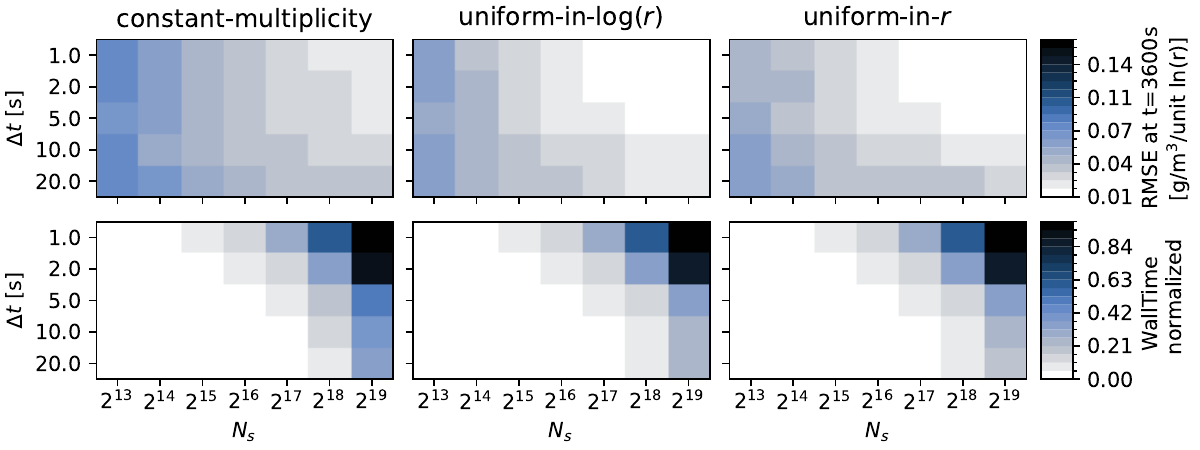}
    \vspace{-1em}
    \caption{Summary of box-model simulations with adaptivity enabled.
    Diagrams constructed as in~Fig.~\ref{fig:heatmaps}, with the same normalization factor for the normalized walltime. (and excluding the deficit which vanishes by design with adaptivity enabled).
    See~Sec.~\ref{sec:box-results} for discussion.
    }
    \label{fig:heatmaps_adaptive}
    \vspace{-1em}
\end{figure*}

\subsection{Setup}

For a box model analysis of our adaptive time-stepping scheme, we use the setup from \citet[][sec.~5.1.4 therein]{Shima_et_al_2009} as shown in Fig. \ref{fig:Shima_2009_deficit}a, tracking the evolution of~the droplet size distribution undergoing collisional growth over one hour (see figure caption for other parameters). 
We use the additive collision kernel proportional to the sum of~the droplet masses $K(x_1,x_2) = b(x_1+x_2)$ for which an analytical solution to the Smoluchowski coagulation equation is known in the case of the exponential initial size distribution
\citep{Safranov_1962,Golovin_1963}.
The coefficient $b$, set here to a value of $1500$~s$^{-1}$, is a collision rate, an early parameterization from calculations on droplet fall speeds in air assuming gravitational coagulation \citep{Golovin_1963}.
While this kernel is not a realistic representation of cloud droplet coagulation, we leverage the analytical solution as a baseline to evaluate the timestepping approach.
Note that, although this kernel admits collisions of same-sized droplets, this is not fully resolved within~the SDM algorithm, as~superdroplets are not tested for self collisions. 

We ran simulations over a range of both $N_s$ and $\Delta t$ to see where the deficit occurs, and how adaptive time-stepping affects the system.
Values within the explored range $2^{12}$ to $2^{19}$ of $N_s$ are larger than $N_s/\text{cell}$ found in LES-type studies in literature, it~is~more comparable to values of system total $N_s$.
We looked at~this for constant-multiplicity, uniform-in-log($r$) and uniform-in-$r$ samplings.
Note that for the given exponential distribution of droplets, uniform-in-$r$ sampling has the largest dynamic range of multiplicity.
We~compared simulation results to the analytical solution at $t = 3600~\mathrm{s}$ using the root-mean-square error (RMSE) of~the droplet size distribution, in~units of~mass density per unit $\log(r)$ (kg m$^{-3}$ / unit $(\ln r)$).
This represents the simulation spread, as~the averaged runs converge to~the solution, except in cases of~very high deficit.
We~bin~the superdroplet mass into 128~bins evenly spaced in~$\log(r)$ from 10~$\upmu$m to~5000~$\upmu$m.
The results are averaged across an ensemble of 30 runs per setting. The initial attributes are sampled deterministically (the same values for all ensemble members) and the ensemble spread comes from Monte Carlo pair selection and event triggering.

\subsection{Result analysis}\label{sec:box-results}
Figure~\ref{fig:heatmaps} summarizes the RMSE and
average collision deficit per time per volume across the tested configurations for the fixed-timestep SDM runs.
As~expected, errors generally decrease with increasing spectral and temporal resolution -- i.e., with more superdroplets and shorter timesteps, respectively.
However, the trend is not strictly monotonic as noted in section~\ref{sec:intro}, especially for the constant multiplicity initialization.
For instance, at $\Delta t = 20~\mathrm{s}$, runs with larger $N_s$ actually have a larger RMSE (in concurrence with a high deficit).
Consistent with previous findings \citep[][]{Dziekan_and_Pawlowska_2017,Unterstrasser_et_al_2020},
runs initialized using uniform-in-log($r$) and uniform-in-($r$) sampling have lower ensemble spread, reflecting the benefit of~resolving the tail of~the droplet size distribution in~collisional growth.
These initializations with a~dynamic range of multiplicity significantly reduce the collision deficit as well, with no~deficit occurring until much longer timesteps, ca.~$\Delta t = 10s$ here.
The magnitude of~the deficit increases with longer timesteps, as well as higher values of $N_s$ (a trend observed more strongly in~the constant-multiplicity runs, although it~is~present in~all attribute sampling strategies).
Although the fact that increasing the~number of~superdroplets increases the deficit may seem counterintuitive, it is reflected in~the probability calculations: the~number of~pairs now experiencing this deficit should increase because there are more pairs being tested ($\lfloor{N_s/2}\rfloor$) with roughly the same pairwise probability (increasing $N_s$ does not affect the average magnitude of pairwise probabilities, because the lower multiplicity is balanced by~the change in~the scaling factor).

Figure~\ref{fig:heatmaps_adaptive} shows the corresponding RMSE for runs using adaptive time-stepping (deficit is not plotted as it vanishes by definition of the adaptive timestepping).
The key observation is that the dependence of~error on~superdroplet count becomes much more robust: increasing $N_s$ now monotonically reduces the error.
The error remains sensitive to the choice of initialization strategy, but greatly reduces the difference at long time steps.
The error is a much stronger function of $N_s$ than it~is~of~timestep, with low numbers of~superdroplets having little to no dependence on $\Delta t$, especially with constant-multiplicity sampling.
This suggests that in~the absence of other processes, there is little to~be gained with a universally small timestep unless the $N_s$ is comparatively large.

In terms of computational cost, adaptive runs scale as expected with both temporal and spectral resolution, but the total cost remains comparable to that of~the base (non-adaptive) timestep configurations.
The cost of adaptive time-stepping is offset by its ability to preserve longer timesteps where safe, leading to similar or even improved runtime performance for a given accuracy level (compare RMSE and walltime between Fig. \ref{fig:heatmaps} and Fig. \ref{fig:heatmaps_adaptive}).

\begin{figure}[t]
    \includegraphics[width=\linewidth]{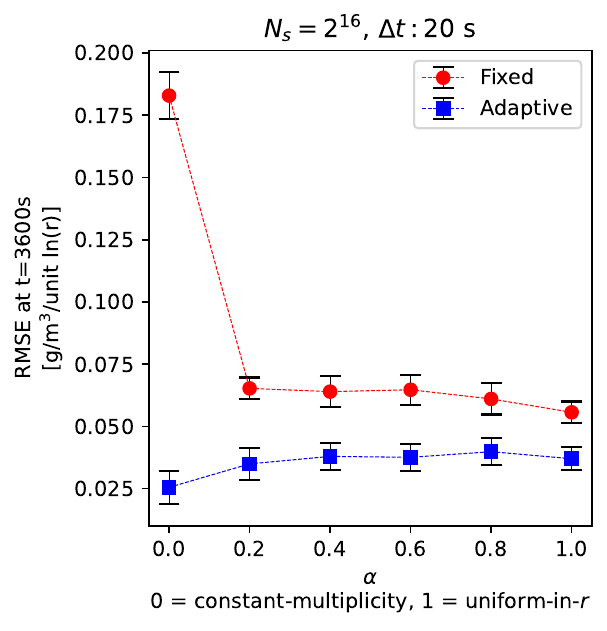}
    \vspace{-2em}
    \caption{
    Dependence of the RMSE on the $\alpha$ parameter 
    for a single setting of $N_\text{s}$ and $\Delta t$.
    The points correspond to ensemble averages over 30 realizations with the error bars indicating $\pm$ one standard deviation.
    Red circles mark results for fixed-timestep runs, blue squares mark runs with adaptivity.
    See Sec.~\ref{sec:box-results}.
    }
    \label{fig:alphaerror}
    \vspace{-1em}
\end{figure}

Figure~\ref{fig:alphaerror} uses the alpha initialization technique described in Section~\ref{sec:alpha} to show selected runs across a continuous range of $\alpha$ values, choosing the end distributions to be constant multiplicity sampling as $\alpha = 0$, and uniform-in-$r$ sampling to be $\alpha = 1$.
Note that this is a different endpoint than 
\citep{Matsushima_et_al_2023} used (uniform-in-log($r$)), because for the exponential distribution this spread is larger.
Adaptive time-stepping runs are shown with a~square symbol.

Enabling the adaptive time-stepping consistently reduces sensitivity to the initialization method.
While the fixed-timestep RMSE varies sharply across $\alpha$, the adaptive runs show flat, stable error profiles.
This suggests that adaptivity not only mitigates the deficit but also improves robustness to sampling choices--especially important in large simulations where the initial sampling resolution is~low with respect to the droplet size distribution or~where multiplicities evolve during runtime.
Figure~\ref{fig:alphaerror} reveals that the constant multiplicity case ($\alpha = 0$) is uniquely prone to elevated deficit and large error.
Even a small increase in~dynamic range ($\alpha > 0$) leads to a notable reduction in~both error and deficit for fixed-timestep SDM.

\begin{figure*}[ht]
    \centering
    \includegraphics[width=\textwidth]{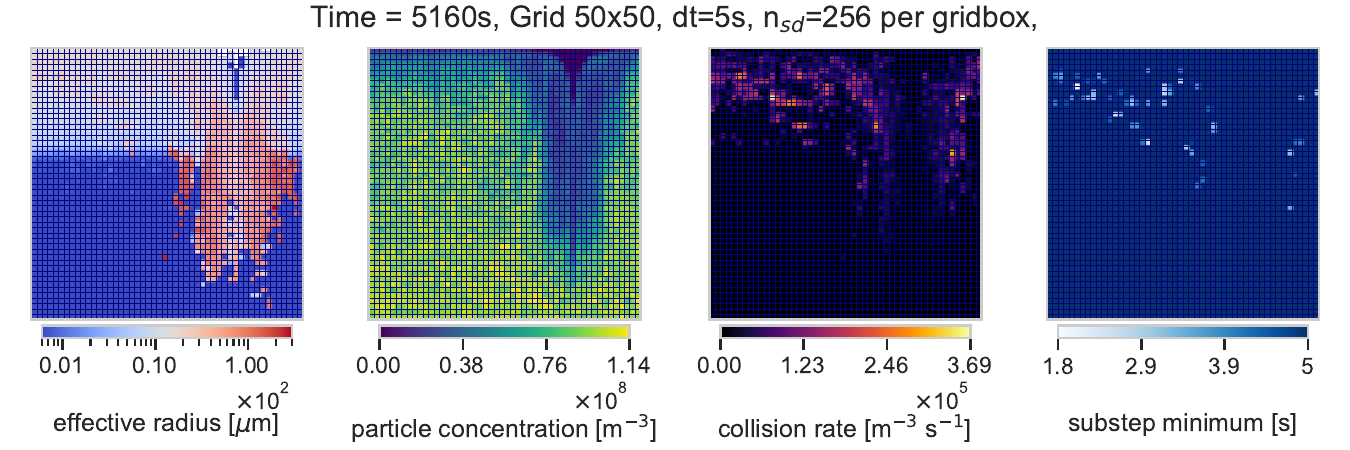}
    \caption{Grid-aggregated particle properties from a single realization of the two-dimensional prescribed-flow simulation.
    The statistics are derived from all particles, regardless of size (incl. aerosol-size, cloud-size, and rain-size regimes). 
    }
    \label{fig:kid2d}
    \vspace{-1em}
\end{figure*}

With adaptive time-stepping enabled, the differences across $\alpha$ values become negligible for cases where it is employed. The adaptive scheme effectively flattens the error landscape, allowing users to~adopt constant multiplicity weighting—valued for its even resolution of~the size distribution—without incurring large penalties from the collision deficit.

Although this shows that the main error in~the constant-multiplicity initialization can be mitigated, having such a difference between the error of~the endpoint and the next alpha value suggests that for when prioritizing spectral accuracy in~the bulk of~the distribution, small alpha values should be used rather than $\alpha=0$, because it will drastically reduce the amount of times an adaptive sub-step is needed.

\section{Flow-coupled simulations}\label{sec:flow}

\subsection{Setup}

To examine the effects of adaptive time-stepping in~a~multi-cell system, we~apply the scheme within a simple two-dimensional prescribed-flow framework  \citep[inspired by][Sec.~3C therein]{Kessler_1969}
 with a~stratocumulus-mimicking setup defined in~\citet{Morrison_and_Grabowski_2007} and employed in multiple subsequent cloud microphysics studies  
as~a~``playground'' for testing algorithms and system behavior with spatial dependence and interactions across grid cells \citep[see][Sec.~4 therein for a list of~over a dozen published studies using the framework]{Arabas_et_al_2025}.

The simulation domain spans $1500~\mathrm{m} \times 1500~\mathrm{m}$, with periodic boundary conditions in~the horizontal ($x$), and vanishing flow velocity at the bottom and the top boundaries.
A prescribed, time-independent air momentum field is defined using a stream function \citep[eq.~1 and 2 of][]{Arabas_et_al_2015}, and a hydrostatic dry-air density profile is~assumed.
Initially, total water mixing ratio and potential temperature fields are populated with a single value everywhere in~the domain ($7.5$~g~kg\textsuperscript{-1} and $289$~K, respectively), leading to supersaturation in~the upper region of~the considered volume.
The two fields are continuously modified by the microphysical processes resolved on the Lagrangian particles (using PySDM), and by the fluid advection resolved on~an~Eulerian grid \citep[here, using the Python just-in-time compiled implementation of~the MPDATA solver -- PyMPDATA,][]{Bartman_et_al_2022_MPDATA}.

Aerosol particles are randomly populated in~the domain (with a uniform distribution) by sampling from a lognormal size distribution and using a single setting of hygroscopicity (proxy for dry aerosol composition).
The supersaturated cloud deck is~initialized to allow droplets to grow to cloud droplet sizes through condensational growth.
To eliminate the influence of~the supersaturation in~the initial condition on the size spectrum of activated cloud droplets, a~spin-up period is employed during which all particles in~the domain are activated and deactivated at least once, with only condensation and advection enabled
\citep[see][section 2.2]{Arabas_et_al_2015}.
After the spin-up period, particle sedimentation and collisions are enabled.
A simple geometric collision kernel is used.
The simulation lasts 90 minutes, out of which the spin-up takes 60 minutes.

We use constant-multiplicity initialization for the superdroplets in~the system.
The evenly weighted sampling is both a common choice in Monte Carlo methods in general, and super-droplet modeling studies as well \citep[e.g. as used in the original SDM paper][]{Shima_et_al_2009}.
As discussed in Sec.~\ref{sec:box-results}, it is also an initialization scheme that maximizes the visibility of~the collision deficit and the effect of~adaptive time-stepping, and avoids masking the effects via initialization strategies that inherently reduce the deficit.
Here we present the results of~this simulation with $n_x=50$ grid cells in both the vertical and horizontal directions, with $N_s\approx256$ superdroplets per cell.
The collision volume extends $1$~m in~the third dimension.
We look at adaptive and fixed time-stepping over a~range of~$\Delta t$ values, with an~ensemble of 40~runs for each of~the three tested $\Delta t$.

\subsection{Result analysis}\label{sec:flow-analysis}

Figure~\ref{fig:kid2d} shows a snapshot from one adaptive run 5160s (26 minutes after collisions are turned on). From left to right, we display: effective radius, particle concentration, collision rate, and minimum allowed collision timestep per grid cell.
The~minimum timestep field reveals where the adaptive substepping has occurred, with timestep values less than $\Delta t = 5$s implying adaptive substepping.
Predictably, the minimum steps are taken where the collision rates are highest, primarily in grid cells in~the middle of~the cloud layer.
Areas where there are large droplets but low number concentration (the rainshaft) do not have as~high collision rates as areas with medium sized droplets and high number concentration.
Practically zero adaptive sub-steps are taken in cells without cloud or rain. Most of~the domain remains at $\Delta t = 5$~s, motivating the adaptive time-stepping scheme rather than suggesting a~need for an overall shorter model time step. 

\begin{figure}[t]
    \centering
    \includegraphics[width=\linewidth]{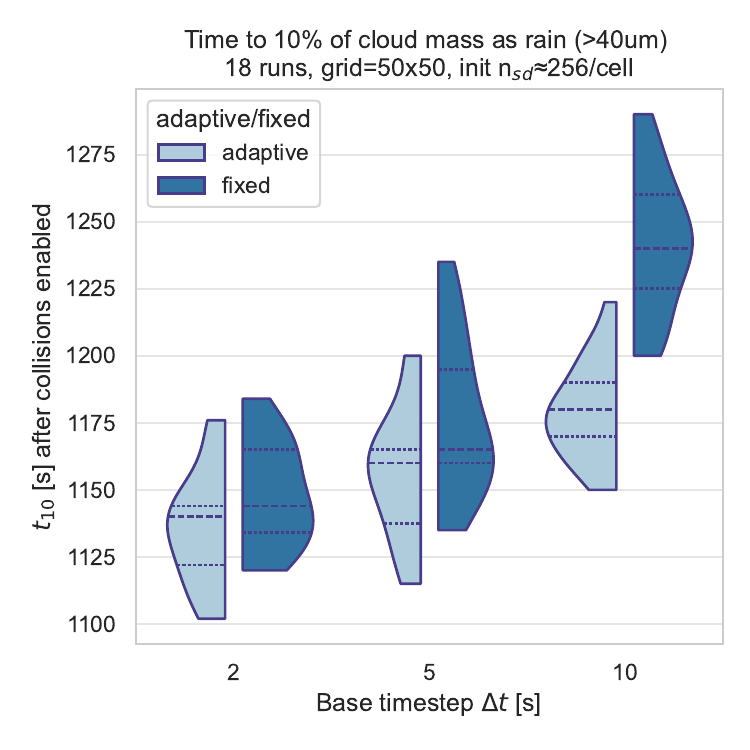}
    \caption{Violin plots showing statistics from an ensemble of 40 realizations showing the $t_{10}$ time (first time $10\%$ of the liquid mass is >40$\upmu m$) for different base timesteps $\Delta t$. The left and right sides of the fiddle show adaptive and non-adaptive runs respectively. Dashed lines denote the median and dotted lines denote interquartile range; edges are cut at the last datapoint.}
    \label{fig:t_10}
    \vspace{-1em}
\end{figure}

We also examine the onset of precipitation using the $t_{10}$ metric \citep[as in][]{Onishi_et_al_2015,Dziekan_and_Pawlowska_2017,Morrison_et_al_2024}, or the first time when $10{\%}$ of~the system liquid mass is contained in raindrop-sized particles ($>40 \upmu$m).
Figure~\ref{fig:t_10} shows that in this configuration, $t_{10}$ is significantly delayed for fixed $\Delta t$ as model timestep increases leading to unrealistically late rain, an effect of~the accumulation of~collision deficit. 
Adaptive runs have $t_{10}$ values much lower than their non-adaptive counterpart and much smaller variation over different $\Delta t$ values, demonstrating how eliminating the collision deficit can aid in accurate rain prediction. 

These results imply that while the average deficit may appear modest at the domain scale, its spatial interaction with the flow field can have a large influence on system-scale properties.
This large of~an~affect is not guaranteed for all configurations (and can be lessened separately through initialization, etc.), the results of~the flow-coupled ensemble runs show that the deficit \textit{can} have a large impact.
Although there are some parameters that can be changed to lower the deficit (lower $\Delta t$, initialization), some factors that create deficit cannot be~changed (number concentration, features of~the underlying droplet size distribution), and some factors are not beneficial to lower resolution (such as~lower $N_s$ or larger grid cells).

\subsection{Multi-threading (incl. GPU) logic}\label{sec:multi-threading}

\begin{figure}[t]
    \centering
    \includegraphics[width=0.5\textwidth]{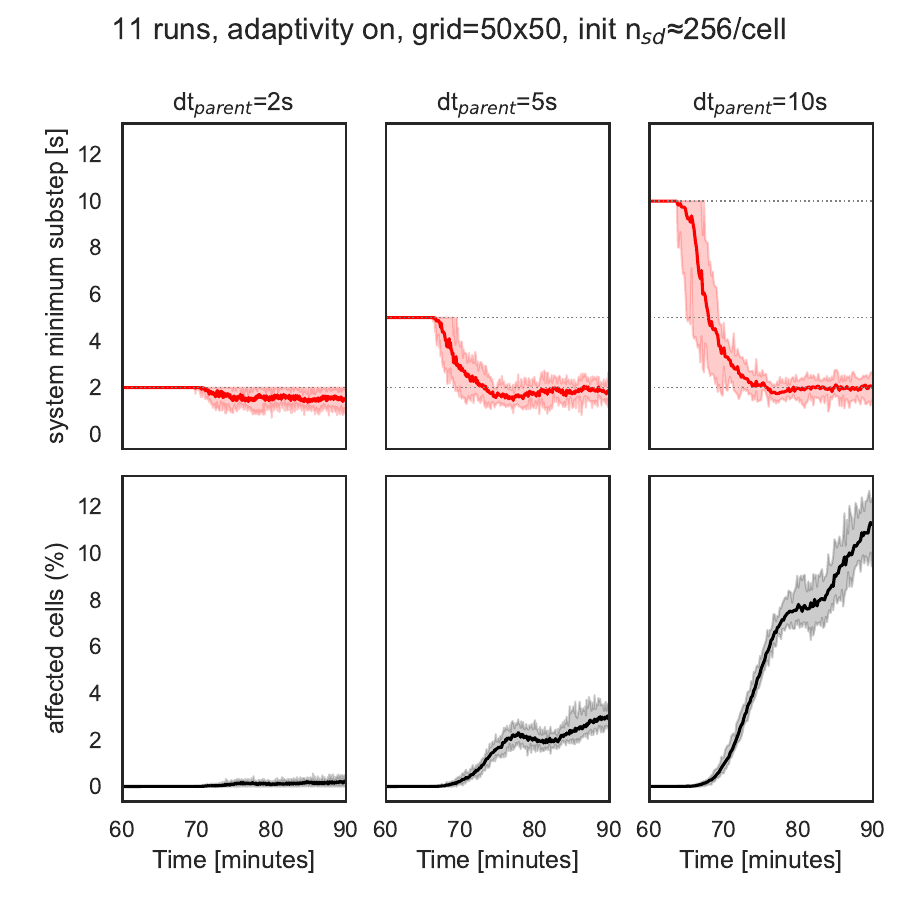}
        \vspace{-2em}
    \caption{Statistics (average and min-max range) from an ensemble of 40 realizations of the flow-coupled simulation setup. The top row shows the minimum adaptive time-step in the whole system over the simulation time, starting after the spin-up period. The bottom row shows the fraction of cells requiring substepping. The columns represent different values of the base timestep~$\Delta t$. See section \ref{sec:flow-analysis} for discussion.}
    \label{fig:fraction_affected}
    \vspace{-1em}
\end{figure}
\begin{figure*}[h!]
    \centering  
    \includegraphics[width=\linewidth]{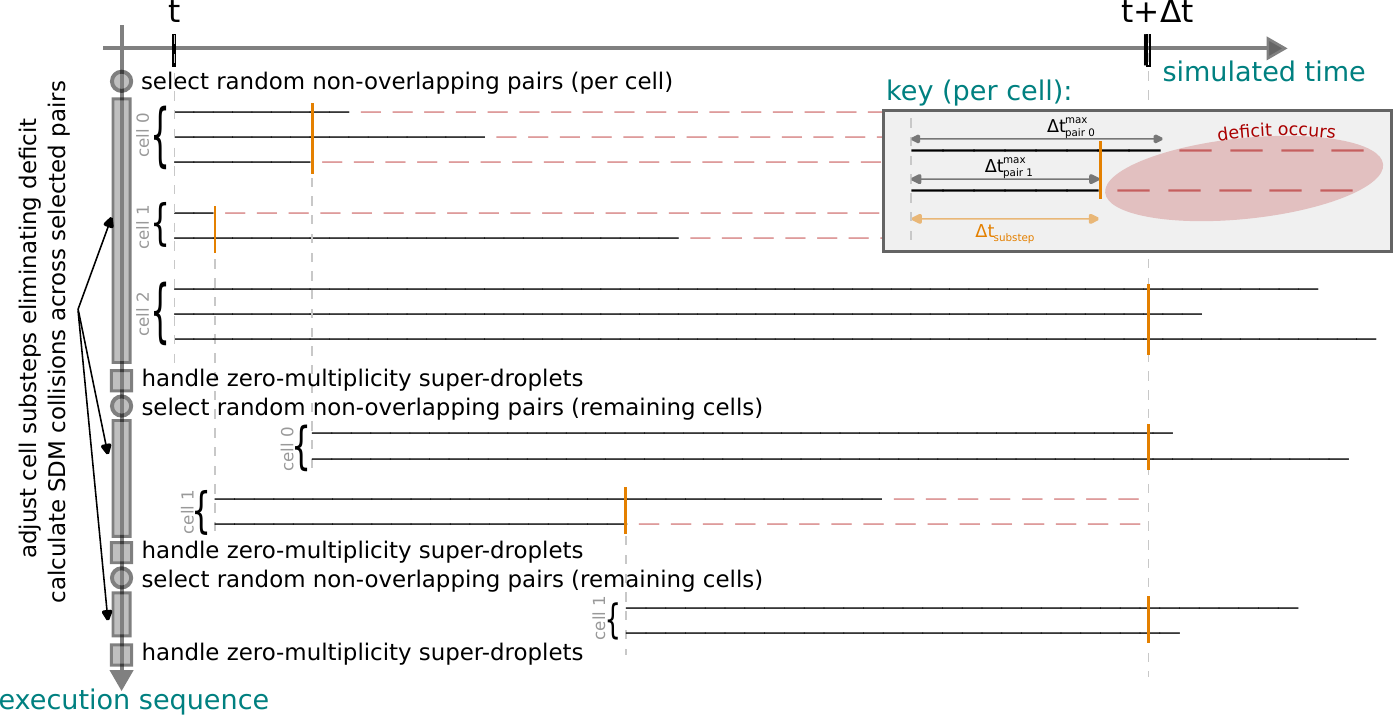}
    \caption{Execution sequence of~the multi-cell adaptive algorithm. Discussion in Sec.~\ref{sec:multi-threading}.}
    \label{fig:multi_cell_alg}
    \vspace{-1em}
\end{figure*}

Analysis of~the 2D flow-coupled simulations described in Section~\ref{sec:flow-analysis} shows that fewer than $10\%$ of~grid cells typically require adaptive substeps, even at coarse base timesteps such as $\Delta t = 10~\mathrm{s}$. 
Figure~\ref{fig:fraction_affected} illustrates this result, showing both the timeseries of~the minimum collision timestep of~the system and the percent of affected grid cells.
This demonstrates that adaptive substepping remains a localized effect, introducing minimal overhead while avoiding the need to globally reduce the timestep.
However, the introduced local timestep variation may lead to global performance bottlenecks if implemented naively in multi-cell models with parallelized environments. 
Although multi-cell implementations of~adaptive time-stepping will largely be model specific, Fig. \ref{fig:multi_cell_alg} provides an example of an execution strategy that supports multi-threading by avoiding idle threads. The collision dynamic is performed across all pairs before taking adaptive substeps, rather than being parallelized over grid-cells. This avoids the bottleneck of all threads being constrained by the system-wide minimum time step.

Figure~\ref{fig:multi_cell_alg} depicts a three cell system completing one base timestep $\Delta t$ in three rounds of substepping.
    Each round consists of:
    (i) selection of random non-overlapping pairs; 
    (ii) adjustment of substep length $\Delta t_\text{substep}$ (per cell) to eliminate deficit;
    (iii) calculation of SDM collisions across all selected pairs (see eq.~\ref{eq:substep});
    (iv) handling of zero-multiplicity super droplets (e.g., removal from the system in~\cite{Shima_et_al_2009} or splitting in \cite{Dziekan_and_Pawlowska_2017}).
While cell 2 completes the full model timestep $\Delta t$ in a single substep, cells~0~and~1 require 2~and~3~substeps respectively, due to~local collision constraints.
In~the first round, cell 0 demonstrates the removal of a zero multiplicity superdroplet, leading to only 2 pairs in~the next round. 
Note that the pairwise rounds facilitate balancing the workload across workers in~a~parallel setup (each worker/thread can handle pairs from multiple cells, and multiple threads can handle pairs from the same cell, the latter being a prerequisite to parallelize the third round in~the diagram).

\section{Conclusions and recommendations}\label{sec:conclusions}

We have discussed results from both box-model and prescribed-flow particle-based cloud microphysics simulations using the SDM algorithm enhanced with an original adaptive time-stepping logic.
The~adaptive scheme is introduced to address the issue of collision deficit, a systematic underestimation of~collision events that arises when superdroplet multiplicities limit the~number of numerically representable collisions within a timestep. 
Using the Safranov-Golovin analytic solution to the Smoluchowski equation for additive kernel and exponential initial size spectrum, we showed that the deficit increases with larger timesteps, higher numbers of~superdroplets (sic!), and lower dynamic range in~multiplicities.
Adaptive time-stepping nullifies this bias without significant computational overhead. 

We leverage the particle attribute sampling scheme proposed in \cite{Matsushima_et_al_2023} to~explore adaptive time-stepping over a continuous range of initial distributions.
We found that constant multiplicity is particularly susceptible to the deficit problem.
While high dynamic range initializations naturally reduce the deficit, even a~small amount of variation in multiplicities (low $\alpha$ values) will lead to~noticeable improvements over \mbox{$\alpha=0$}.
Adaptivity reduces the sensitivity to the sampling scheme, allowing users to employ initialization methods that are optimal for other processes (e.g.,~condensation).
In two-dimensional flow-coupled simulations, we~found that there are situations where the collision deficit can have a~strong influence on~bulk properties, larger than previously observed in multi-cell systems \citep{Unterstrasser_et_al_2020}. 
Although the deficit affects only a~small fraction of~the grid cells with high collision rates, these local hotspots can delay precipitation onset if~left uncorrected.
Adaptive time-stepping resolves this by targeting only the regions where the deficit occurs, preserving accuracy while minimizing the computational cost.
While uniformly reducing the~timestep can mitigate the deficit or even have runs that never experience this limit, adaptive time-stepping is~the \textit{only} way to~systematically eliminate it.
Based on~these findings, we~strongly recommend implementing adaptive time-stepping as a default in~SDM-based models (the adaptivity logic applies as~well to the collisional breakup extension of SDM developed in~\citeauthor{de_Jong_et_al_2023_GMD}, \citeyear{de_Jong_et_al_2023_GMD}).
It enables robust and accurate simulations across a~wide range of configurations, reduces sensitivity to user-defined parameters, and ensures physical consistency without compromising performance.
This~makes it especially valuable in~operational or large-scale modeling systems where timestep constraints and initialization choices are dictated by broader modeling frameworks.

\appendix

\begin{figure*}[ht!]

    \includegraphics[width=\textwidth]{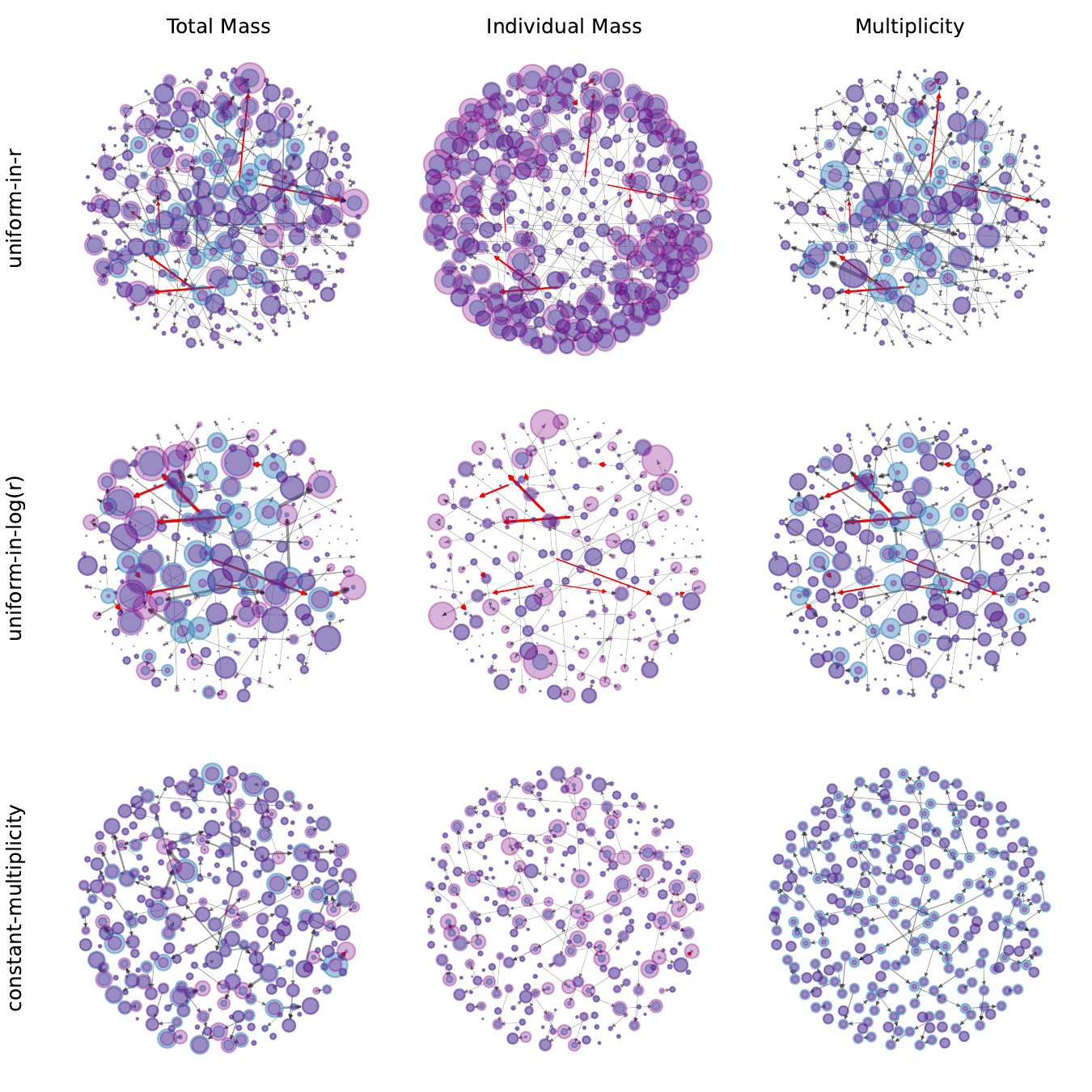}
    \caption{\label{fig:network}
    Directional mass flows between superdroplets over two simulation steps collapsed into a single network graph. Each row of panels utilizes a different attribute sampling method (see section \ref{sec:alpha}). The sizes of the nodes in each column are proportional to superdroplet total mass, individual mass, and multiplicity, with 
    initial and the final weighting depicted with blue and purple shades, respectively. 
    The~black arrows represent collision events with width scaling proportional to the transfer of total mass, individual mass, and multiplicity respectively.
    Red arrows depict the collision deficit.
    See~Appendix~\ref{sec:network} for discussion.
    }
\end{figure*}

\renewcommand{\thesection}{Appendix \Alph{section}}

\section{Collision interaction networks}\label{sec:network}

Figure~\ref{fig:network} depicts two timesteps of a 
box-model, Golovin-Safranov kernel simulation with $2^8$ super droplets and $\Delta t=150$~s (all other parameters ~in Fig.~\ref{fig:Shima_2009_deficit} and Fig.~\ref{fig:heatmaps}). The large $\Delta t$ is chosen to amplify the deficit. Visualization created with the NetworkX package \citep{Hagberg_et_al_2008}.
The collision events are depicted as a network of directional collisional interactions.
The visualizations are done separately for the total mass, individual mass multiplicity (columns) and for three considered attribute initialization methods (rows).

The deficit is depicted with red arrows, and is~most prominent in the uniform-in-log($r$) case, less prominent in the uniform-in-$r$, in line with results presented in Fig.~\ref{fig:heatmaps}.
The least visible deficit is~for the constant-multiplicity case, which corresponds to what would be expected by extrapolating the slope of deficit profile to the low number of superdroplets (outside of the range presented in Fig.~\ref{fig:heatmaps}). 

The nodes are rendered with semi-transparent blue and purple shades depicting initial and final states.
Note that due to both individual masses and multiplicities changing over time, the final total mass represented by a superdroplet can be either smaller or larger than the total initial mass (left column plots), while individual masses are monotonically increasing (middle column) and multiplicities are monotonically decreasing (right column).

As highlighted in Sec.~\ref{sec:box-results}, attribute sampling strategies with a dynamic range of multiplicities increase the number of events for a given number of~superdroplets.
This can be seen in Fig. \ref{fig:network} with the number of black lines in the plots, with the highest amount in the first row and the least in the last row (e.g., note the number of line crossings). 

The network layout optimizes the positions of the superdroplets by placing those with the least number of collision interactions outwards \citep[degree dependent repulsive force, see][]{Jacomy_et_al_2014}. 
Although there are only two time steps, this is enough to see that the low-mass/high-multiplicity superdroplets in the uniform-in-r sampling experience more events than the high-mass/low-multiplicity droplets, and that this is distint for the sampling method.

\clearpage
\subsubsection*{Acknowledgements}

Thank you to Obin Sturm, Simon Unterstrasser and Shin-ichiro Shima for helpful discussions and advice throughout the duration of the project.
EW and SA acknowledge support from the Polish National Science Centre (grant no.~2020/39/D/ST10/01220). 
SA~acknowledges support from the AGH Excellence Initiative -- Research University (Grant IDUB 9056).

\subsubsection*{Author Contributions}

The concept of the deficit, the adaptivity scheme and the threading logic were introduced by PB and implemented in~PySDM in his MSc project (Jagiellonian University, 2020, under the mentorship of~SA). 
Based on it, an independent Julia implementation of the adaptivity scheme in the Droplets.jl package was developed by EW (under the mentorship of ALI).
The $\alpha$-sampling module for PySDM was developed by EW.
The simulations and analyses presented and discussed in the paper were conceptualized by EW and SA, and carried out by EW during an Erasmus+ year at~AGH University of~Krakow.
The paper and simulation figures were composed by EW with input from all authors.

\subsubsection*{References}
\renewcommand*{\bibfont}{\small}
\printbibliography[heading=none]

\end{document}